\documentclass[doublecol]{epl2} 
\pdfoutput=1
\usepackage{amsmath}
\usepackage{graphicx}
\usepackage{amssymb}
\usepackage{bm}
\title{Transport properties of graphene with one-dimensional charge defects}
\shorttitle{Transport properties of graphene with one-dimensional charge defects} 

\author{
Aires Ferreira\inst{1} 
\and
Xiangfan Xu\inst{2}
\and
Chang-Lin Tan\inst{3}
\and 
Su-Kang Bae\inst{3}
\and
N. M. R. Peres\inst{1}
\and
Byung-Hee Hong\inst{3, 4}
\and
Barbaros \"Ozyilmaz \inst{2,5}
\and
A. H. Castro Neto \inst{6}
}
\shortauthor{Aires Ferreira \etal}

\institute{                    
 \inst{1} Department of Physics and Center of Physics, University of Minho, P-4710-057, Braga, Portugal, EU\\
 \inst{2} Department of Physics, 2 Science Drive 3, National University of Singapore, Singapore 117542\\
 \inst{3}SKKU Advanced Institute of Nanotechnology (SAINT) and Center for Human Interface Nano Technology (HINT), Sungkyunkwan University, Suwon 440-746, Korea\\
 \inst{4}Department of Chemistry, Sungkyunkwan University, Suwon 440-746, Korea\\
 \inst{5}NanoCore, 4 Engineering Drive 3, National University of Singapore, Singapore 117576\\
 \inst{6}Department of Physics, Boston University, 590 Commonwealth Avenue, Boston, MA 02215, USA
}

\abstract{We study the effect of extended charge defects in electronic transport
properties of graphene. Extended defects are ubiquitous in chemically and 
epitaxially grown graphene samples due to internal strains associated with the
lattice mismatch. We show that at low energies these defects interact quite 
strongly with the 2D Dirac fermions and have an important effect in the DC-conductivity
of these materials.}

\begin{document}
\maketitle
\section{Introduction}
Graphene crystals isolated by the exfoliation method \cite{Novoselov}
are high quality films, with high mobilities even on a SiO$_2$ substrate. While exfoliation
works well for the study of the fundamental physical properties of graphene \cite{review}
it is not a scalable process useful for technological applications \cite{rise}. At the present
time, the most promising scalable growth methods of graphene  films are either based on
epitaxial growth on SiC \cite{deheer} or on chemical vapor deposition (CVD) of graphene
on metal surfaces \cite{Singapore_experiment}.

Graphene growth on crystal substrates, independently of the process, is subject to strain
due to lattice mismatch between graphene and the substrate. The strong sigma bonding 
between the carbon atoms makes the graphene lattice very stiff (the spring constant is of
order of 50 eV/\AA$^2$) and hence in-plane deformations are energetically costly.  
Because graphene is the ultimate example of a 2D film, the strain can be release by two
main mechanisms, namely, either by going out of the plane or by the reconstruction of the
chemical bonds. 

By exploring the third dimension the graphene film pays the energetic price of bending 
(the bending rigidity is of order of 1 eV) and the loss of interaction energy with the substrate
(which is usually a mix of covalent bonding and van der Waals interaction). In certain cases,
the energetic price of forming wrinkles  and blisters in the graphene film is smaller than the
price of creating structural defects such as pentagons, heptagons, and octagons. STM studies
have shown that epitaxial graphene grown on 6H-SiC actually bends and buckles as a result
of the compressive strain \cite{sun}. The same effect is observed in samples grown by CVD
on Cu \cite{Graphene_copper}.  In other cases, when the the interaction between graphene
and the substrate is strong, it is energetically preferable to reconstruct the bonds with the 
formation of lines of defects. This is what happens, for instance, in graphene grown by CVD
on a Ni surface \cite{metallic_wire}. The surface of the film reveals the presence of extended 
one-dimensional  defects. These defects are lines of periodic cells made of two pentagons 
followed by one octagon.   

The general conclusion is that the intrinsic 2D nature of graphene makes the presence of 
one-dimensional extended defects in artificially grown graphene samples a norm. The 
formation of extended structural defects in graphene has strong consequences for the 
electronic properties. On the one hand, localized bending and strain can lead to the appearance of strong ``pseudo-magnetic'' fields that create localized Landau levels. This is the main idea behind the concept of strain engineering \cite{vitor}. On the other hand pentagons, heptagons, and octagons, act as donors and/or acceptors relative to the normal benzene ring configuration, and hence also leads to charge localization \cite{yazyev}. All these mechanisms are examples of the more general concept of {\it self-doping} in sp$^2$ bonded carbon \cite{Peres}. The existence of the localized states at the defect line lead to transfer of charge from the bulk of graphene to the defect --- charge accumulates at the extended defect, creating charged lines. Such type of scattering lines can be the limiting scattering mechanism of the electronic mobility in these graphene films. Furthermore, experimental studies show that these extended defect lines intercept each other at random angles, forming irregular polygons with edges showing stochastic distribution of length. A study of the electronic scattering in chemically produced graphene needs to take into account such types of random distributions in order to correctly describe the effect of the extended charge defects in DC-transport.
 
This Letter is organized as follows; we begin by outlining the experimental procedure used to produce the CVD graphene samples and briefly characterize the extended defects seen in the microscopy studies. Then, the main result is presented and tested against DC-conductivity data for several CVD graphene samples. In the remainder we discuss in detail the model of extended charge defects, the effective scattering potential and its contribution to the semi-classical conductivity. Last, finer points of our calculation, such as the effect of disorder in the length distribution of the defects, and the electron-hole asymmetry of electronic cross-sections  due these defects are discussed in separate sections.
\begin{figure}
\begin{centering}
\includegraphics[width=0.7\columnwidth]{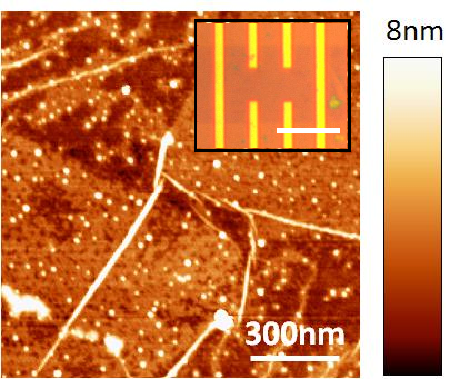} 
\par\end{centering}
\caption{\label{fig_1a}
Tapping mode atomic force microscope (AFM) image of a CVD graphene film transferred on SiO$_2$. Different thermal expansion of the Cu foil and the graphene sheet result in the formation of a few nm high ripples. Locally cracks can form during the transfer process and occasionally one is left with PMMA residues. Inset: Hall bar device used to measure the conductivity of CVD grown graphene (scale bar is 5 $\mu$m). Representative data is given in Figs. \ref{fig:fig2} and~\ref{fig:fig2b}.}
\end{figure}
\begin{figure}
\begin{centering}
\includegraphics[width=0.9\columnwidth]{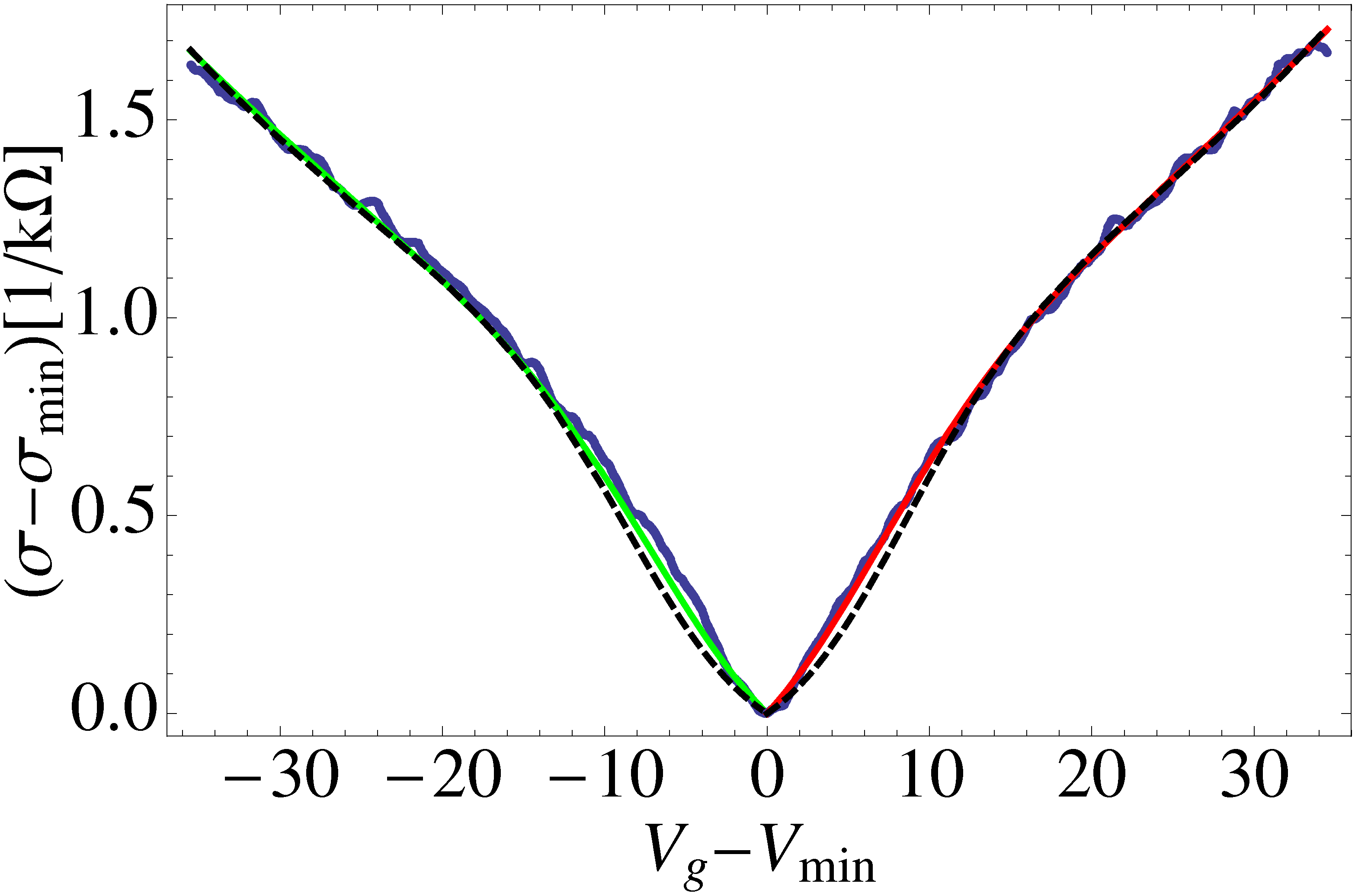} 
\par\end{centering}
\caption{\label{fig:fig2}
Conductivity measured in CVD synthesized graphene at T=3.5K (data points shown in blue) \cite{Singapore_experiment} is fitted to Eq.~\ref{eq:DC_Conductivity} as function of $V=V_{g}-V_{\textrm{min}}$ (dashed line). The optimal parameters are found to be $W\simeq14.6\textrm{nm}$; $\gamma \simeq2.14\times10^{11}\textrm{\ensuremath{\textrm{cm}^{-2}}}$($V\ge0$) and $\gamma \simeq2.28\times10^{11}\textrm{\ensuremath{\textrm{cm}^{-2}}}$($V<0$). The inclusion of midgap states (i.e. resonant scatterers) is shown in thick lines {[}red ($V\ge0$) and green ($V<0$){]} and modifies $\gamma$ to $1.15\times10^{11}\textrm{\ensuremath{\textrm{m}^{-2}}}$ ($V\ge0$) and $1.27\times10^{11}\textrm{\ensuremath{\textrm{m}^{-2}}}$ ($V<0$). The midgap parameters are $n_{s}=1.6\times10^{11}\textrm{cm}^{-2}$ and $R=2a_{0}$, with $a_{0}=0.14\textrm{nm}$; the experimental data was shifted as to have a minimum of zero at
the Dirac point ($V_{\textrm{min}}\simeq$5.5V and $\sigma_{\textrm{min}}\simeq$0.1$e^2/h$). Fits to other CVD samples are shown in Fig.~\ref{fig:fig2b}.}
\end{figure}
\section{Outline}
The theory described below has been used to interpret the transport data of graphene grown using the roll-to-roll method \cite{rolltoroll}. An AFM image of a CVD graphene film transferred to SiO$_2$ is shown in Fig.~\ref{fig_1a}; extended line defects, few nanometers long, can be seen.
Sample preparation and measurement were performed using standard methods: graphene samples are grown on Cu substrates by chemical vapor deposition (CVD) \cite{rolltoroll}. To characterize CVD graphene samples, standard Hall bars are patterned by e-beam lithography, followed by thermal evaporation of Cr/Au (5/25 nm). An additional e-beam lithography step followed by O$_2$ plasma etching is performed to define Hall bar device as shown in the inset of Fig. \ref{fig_1a}. Measurements  are performed as a function of temperature down to 3.5 K using standard lock-in amplifier techniques. 

The central result of this work is an expression for the semi-classical DC-conductivity of graphene due to extended charge defects,
\begin{equation}
\sigma_{l}=\frac{16e^{2}}{h}\frac{\pi k_{F}^{2}}{n_{l}}\left(\frac{e}{q_{l}}\right)^{2}G(k_{F}W)\,,
\label{eq:DC_Conductivity}
\end{equation}
where $n_{l}$ is a finite density of effective extended defects (see next section), made of lines with charge $q_{l}$ and $G^{-1}(x)=\int\,\, d\theta(1-\cos\theta)\left(\pi-\theta\right)^{2} \left[5+4\cos\left(x-x\cos\theta\right)\right] $. For sake of simplicity, in deriving the latter equation, we have considered a graphene structure constant $\alpha=1/2$. The Fermi momentum $k_{F}$ relates to the electronic carrier density $n_{c}$ according to $k_{F}=\sqrt{\pi n_{c}}$, which can be controlled by the application of a back-gate voltage, $V_{g}$, after transferring the graphene sheet to a dielectric substrate, typically silicon oxide, $300\mbox{nm}$ thickness, for which one has $n_{c}=7.2\times10^{14}V_{g}$ (SI units). To test our results, we used experimental data of conductivity measurements in several graphene samples grown by CVD on Cu, using the roll-to-roll method \cite{rolltoroll}. The fitting parameters in our theory are $W$ and $\gamma \equiv n_{l}q_{l}^{2}/e^{2}$. The former has units of length and is roughly the mesh size originated by the intersection of defect lines, whereas the latter measures the scattering strength of the extended defects. Figure~\ref{fig:fig2} shows a high-quality fit to the CVD data. For moderate to high carrier densities, we find that Eq.~\ref{eq:DC_Conductivity} fits perfectly well the data, indicating that linear charged lines formed during the CVD growth process can play a key role in DC-transport at finite electronic densities. The line separation was found to lie within the range 10-15 nm in all samples. We also found that the inclusion of strong (resonant) short-range disorder improves the agreement with the experimental data for lower carrier densities, more precisely,  $|V_g-V_{\textrm{min}}| \lesssim 10$V. (Recall that resonant scattering has recently emerged as one of the main mechanisms limiting DC-transport of non-suspended graphene \cite{ReviewPeres,RS}.) In what follows, we give a detailed description of our model of extended charge defects in graphene.
\section{Theory}
\begin{figure}
\begin{centering}
\includegraphics[width=0.7\columnwidth]{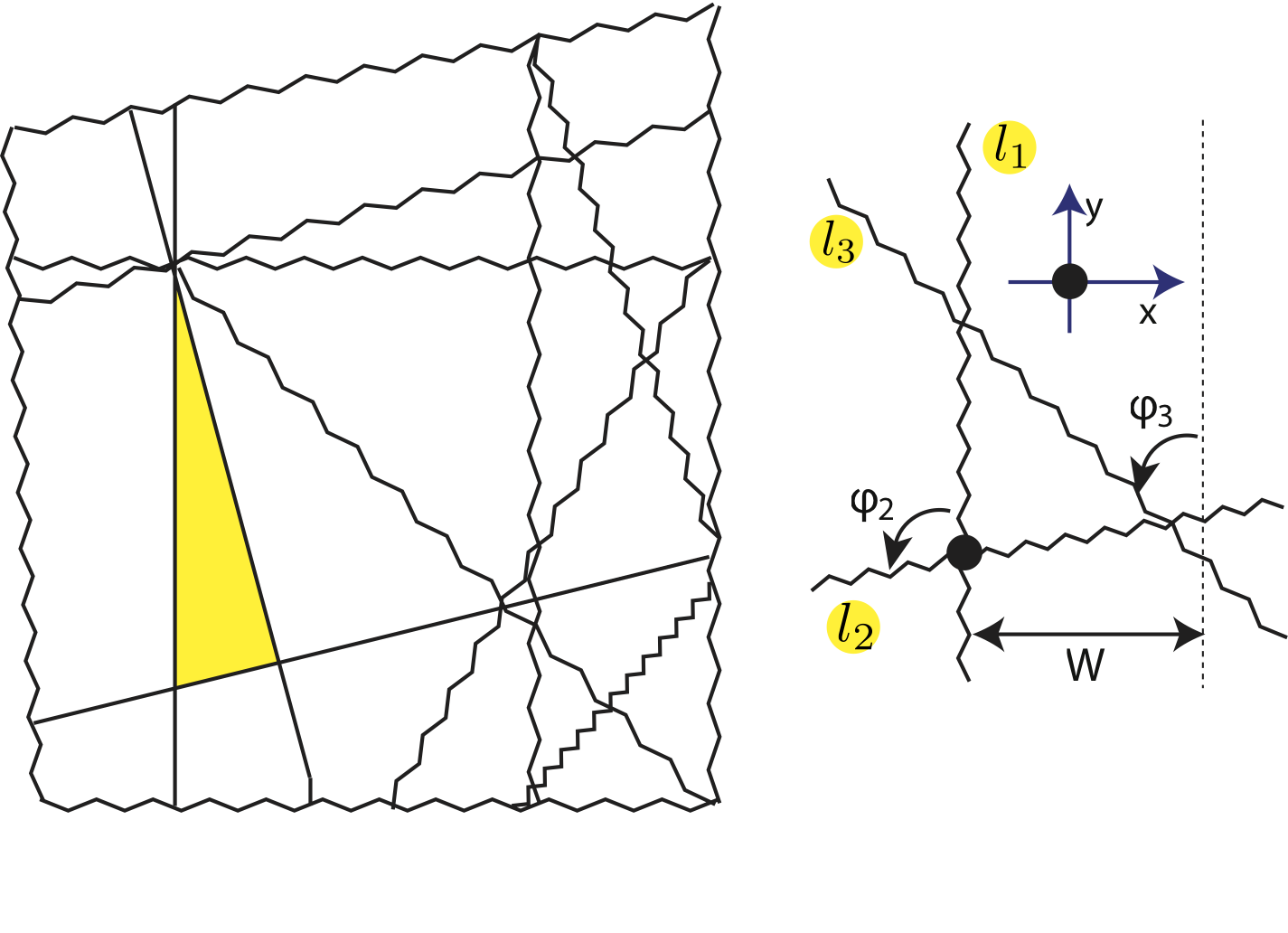} 
\par\end{centering}
\caption{\label{fig:fig1}Left --- Schematic of a small area of CVD graphene with charged lines with several length; the area (in yellow) encompassed by $3$ lines defines the intersection zone of the effective extended charged defect. Right --- The effective extended defect is built, from three lines initially at $x=0$, translating one line (e.g. $l_{3}$) along the $x$-axis by $W$ and rotating the lines by $\varphi_{i}$ about the $z$-axis. In the picture $\varphi_{1}=0$ to make it easy to visualize; the origin of the coordinate system is represented as a black dot.}
\end{figure}
We start by characterizing the scattering potential created by a single extended charge defect. We take the charged defect to be an infinite line along the $y$ axis, with linear charge density $\lambda_{l}$, embedded in graphene, which is lying in the $xy-$plane. Its 3D charge density has the form $\rho_{\textrm{line}}(r)=\lambda_{l}\delta(x)\delta(z)$. Basic electrostatics predicts that single line of charge in vacuum produces a logarithmic potential in space. Clearly, embedded in a metal, the potential is modified by screening effects. The screening can be taken into account within the Thomas-Fermi (TF) approximation \cite{Ziman}. The TF assumes that the local electronic charge  density, $\rho(\mathbf{r})$, changes due to the effective potential, $\phi_{\textrm{eff}}(\mathbf{r})$, created by the extended defect according to: 
\begin{equation}
\delta\rho(\mathbf{r})\simeq-e\rho(E_{F})e\phi_{\textrm{eff}}(\mathbf{r}),
\label{eq:epl1}
\end{equation}
where $\rho(E_{F})=2k_{F}/(\pi\hbar v_{F})$ is the bare density of states per unit of area (spin and valley degeneracies included) and $k_{F}$ ($v_{F}$) is the Fermi momentum (velocity). 

The effective potential, $\phi_{\textrm{eff}}$, is determined by Poisson's equation: $\nabla^{2}\phi_{\textrm{eff}}=-\left(\rho_{\textrm{line}}+\delta\rho\right)/\epsilon_{d}$, that is, 
\begin{equation}
\epsilon_{d}\nabla^{2}\phi_{\textrm{eff}}(x,z)=\left[\frac{2e^{2}k_{F}}{\pi\hbar v_{F}}\phi_{\textrm{eff}}(x)-\lambda_{l}\delta(x)\right]\delta(z),
\label{eq:epl2}
\end{equation}
where $\epsilon_{d}$ is the dielectric constant of the medium [notice that $\delta\rho=-e^ 2\rho(E_{F})\phi_{\textrm{eff}}(x,z=0)$]. This equation can be solved by Fourier transform followed by an integration over the $z$ coordinate. The form of the potential in momentum space is
\begin{equation}
\tilde{\phi}_{\textrm{eff}}(q_{x})=\frac{\lambda_{l}}{2\epsilon_{d}}\frac{1}{|q_{x}|+q_{TF}},
\label{eq:Potential_Infinite_Line}
\end{equation}
where $q_{TF}\equiv4\alpha k_{F}$ is the TF wave vector, with the effective graphene's structure function given by  $\alpha\equiv e^{2}/(4\pi\epsilon_{d} \hbar v_{F})$. We note that $\phi_{\textrm{eff}}(x)$ shows a logarithmic divergence at the origin and asymptotic behavior ($q_{TF}x \gg 1$) given by 
\begin{equation}
\phi_{\textrm{eff}} (x)\rightarrow\lambda_{l}/\left[2\epsilon_{d}\pi(q_{TF}x)^{2}\right].
\label{epl2b}
\end{equation}
In contrast with the screened Coulomb potential created by a point charged impurity, the potential of a charged defect line does not decay exponentially away from the scattering center, and hence can lead to a strong effect on electronic transport, 
as we show below.

Atomic force microscopy of small areas of graphene ($\sim$0.1$\mu$m$^{2}$) shows extended defects with several lengths and orientations \cite{Singapore_experiment}.
We model these extended defects as straight lines that intersect forming polygons (see Fig.~\ref{fig:fig1}). The electronic scattering is determined by the polygons formed by the defect lines (see later). The network of defects is built from a number of base lines (labeled $l_{i}$), initially lying along the $y$-axis, in two steps: translating each line $l_{i}$ by a vector
$\mathbf{R}_{i}=(x_{i},y_{i}$), and finally rotating them about the $z$-axis by random angles $\{\varphi_{i}\}$. Throughout the paper we assume that the size $L$ of these lines is much larger than other length scales in the problem. For $N$ base lines the screened potential reads:
\begin{equation}
\tilde{\phi}_{\textrm{eff}}^{N}(\mathbf{q})=\sum_{i=1}^{N} \frac{q_{i}}{2\epsilon_{d}}
\frac{e^{-i\mathbf{q}\cdot\mathbf{R}_{i}}}{|Q(\varphi_{i})|+q_{TF}},
\label{eq:epl3}
\end{equation}
where $Q(\varphi_{i})=q_{x}\cos\varphi_{i}-q_{y}\sin\varphi_{i}$ is the projection of the wave vector $\mathbf{q}$ onto the direction perpendicular to the line $l_{i}$, defining the direction of momentum transfer in an electron-defect scattering event, and $q_{i}\equiv\lambda_{l}L_{i}$
is the charge of line $l_{i}$. [To obtain this result, we approximated the 2D electrostatic potential of a single line (e.g. oriented along the $y$ axis) using $\tilde{\phi}_{\textrm{eff}}(\mathbf{q})=\int dy \phi_{\textrm{eff}}(q_{x})\theta(L/2-|y|)$, where $\theta(y)$ is the Heaviside function and $\phi_{\textrm{eff}}(x)$ is the potential of an infinite charged line embedded in graphene (Eq.~\ref{eq:Potential_Infinite_Line}).] 
\subsection{Model of the extended defect}
\begin{figure}
\begin{centering}
\includegraphics[width=0.6\columnwidth]{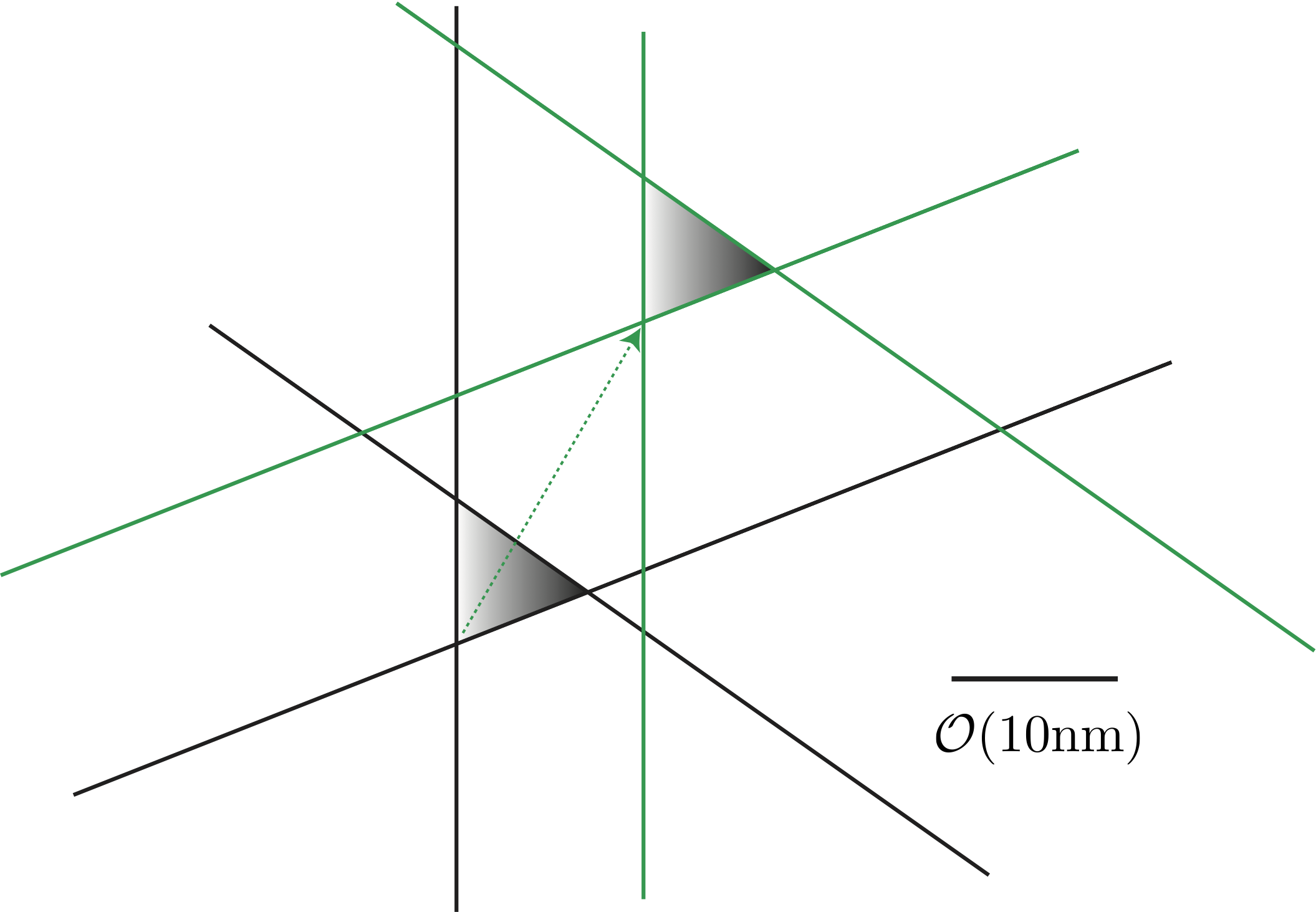} 
\par\end{centering}
\caption{\label{fig:fig3}An effective extended defect with triangular intersection (three lines in black) is replicated and translated according to the green arrow, producing a mesh containing several types of irregular polygons.}
\end{figure}

Transport through nano-electronic graphene devices with extended charge defects is tackled here from the point of view of a single effective defect ``cell'', from which a prototype network of charged lines can be reproduced. The effective extended defect is made of three lines, with equal lengths, which intersect forming a triangle (see Fig.~\ref{fig:fig1}); choosing $\mathbf{R}_{1(2)}=(0,0)$ and $\mathbf{R}_{3}=(W,0)$, introduces a new length scale, $W$, the amount of translation of $l_{3}$, here referred to as line separation. Its scattering potential (denoted by $\phi_{\bigtriangleup,\textrm{eff}}$) is given by Eq.~(\ref{eq:epl3}) setting $N=3$. Figure~\ref{fig:fig3} shows how many such defects give rise to all sort of polygons. 

The large-distance behavior of the potential due to a charged line in graphene (see Eq.~\ref{eq:Potential_Infinite_Line}) renders the Born series particularly suitable to compute scattering amplitudes. In the first Born approximation (FBA), the elastic scattering amplitude for massless Dirac fermions in 2D reads \cite{Novikov}
\begin{equation}
f(\theta)=\frac{\Xi(\theta)}{\hbar v_{F}} \sqrt{\frac{|\mathbf{p}|}{8\pi}}
e\langle\tilde\phi_{\bigtriangleup,\textrm{eff}}(\mathbf{q})\rangle,
\label{eq:epl4}
\end{equation}
where $\mathbf{p}$ ($|\mathbf{p}|=k_{F}$) is the wave vector of the incident electron (we choose $\mathbf{p}=|\mathbf{p}|\mathbf{e}_{x}$), $\mathbf{q}=\mathbf{p}^{\prime}-\mathbf{p}$ the transferred wave vector ($\mathbf{p}^{\prime}$ stands for the {}``out'' wave vector), $\theta=\angle(\mathbf{p}^{\prime},\mathbf{p})$ is the scattering angle and $\langle\phi_{\bigtriangleup,\textrm{eff}}(\mathbf{q})\rangle$ is the scattering potential conveniently averaged as to include rotational disorder (disorder in $W$ will also be considered);  the factor
$\Xi(\theta)=1+e^{i\theta}$ originates from graphene's Berry phase precluding electrons from back-scatter \cite{Ando}. 
The main result (\ref{eq:DC_Conductivity}) is obtained within the Boltzmann approach using the FBA,
\begin{equation}
\sigma_l=(1/2)e^{2}v_{F}^{2}\rho(E_{F})\tau_l,
\label{eq:CondSC}
\end{equation}
where the relaxation time is given by $1/\tau_{l}=n_{l}v_{F}\int d\theta(1-\cos\theta)|f(\theta)|^{2}$
\cite{Ziman}. 
\subsection{Rotational disorder}
\begin{figure}
\begin{centering}
\includegraphics[width=0.8\columnwidth]{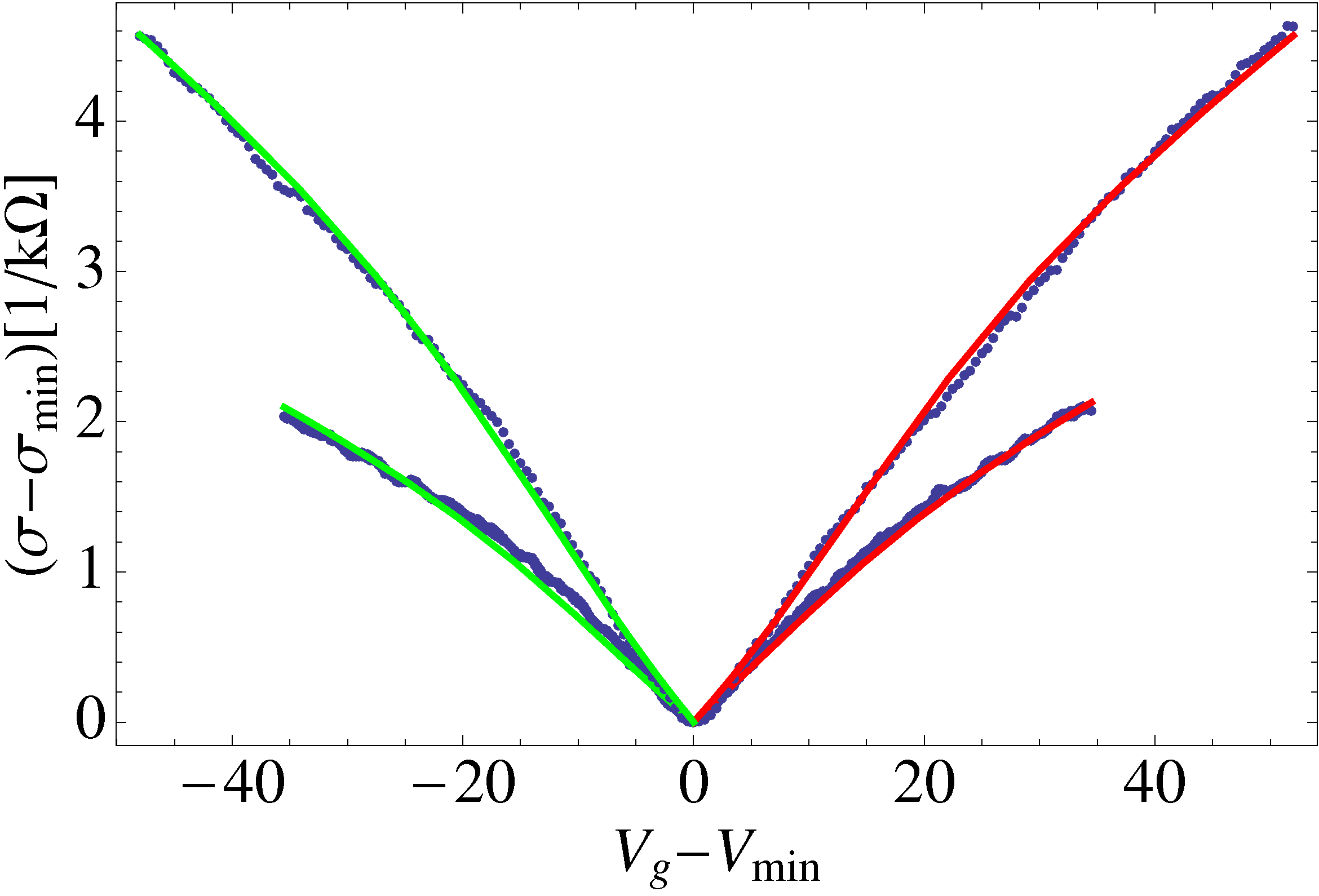} 
\par\end{centering}
\caption{\label{fig:fig2b}
Conductivity measured in CVD synthesized graphene at T=3.5K for two samples with different mobilities (blue dots) and respective fits to the semi-classical calculation with resonant scatterers contribution included. Sample with higher (lower) mobility: $W\simeq10.0(11.5)\textrm{nm}$; $\gamma \simeq5.38(4.84)\times10^{10}\textrm{\ensuremath{\textrm{cm}^{-2}}}$($V\ge0$) and $\gamma \simeq4.63(5.56)\times10^{10}\textrm{\ensuremath{\textrm{cm}^{-2}}}$($V<0$). The midgap parameters are $n_{s}=1.5(3.0)\times10^{11}\textrm{cm}^{-2}$ and $R=a_{0}$, with $a_{0}=0.14\textrm{nm}$; the experimental data was shifted as to have a minimum of zero at the Dirac point: $V_{\textrm{min}}\simeq$8.0(5.5)V and $\sigma_{\textrm{min}}\simeq$0.2 (0.1)$e^2/h$.
}
\end{figure}

Within the model depicted in Fig.~\ref{fig:fig1}, rotational disorder can be taken into account by averaging the scattering potential $\phi_{\bigtriangleup,\textrm{eff}}$ over the angles $\{\varphi_{i}\}$. The scattering potential due to a single $\triangle$ extended charge defect was averaged using a uniform random distribution of angles:
\begin{equation}
\langle\tilde\phi_{\bigtriangleup,\textrm{eff}}(\mathbf{q}) \rangle\equiv\pi^{-3}\int_{0}^{\pi}\prod_{i=1}^{3}d \varphi_{i}\tilde\phi_{\bigtriangleup,\textrm{eff}}(\mathbf{q}).
\label{eq:epl5}
\end{equation}
The influence of the particular lines orientation on transport depends strongly on the screening length. For graphene on top of SiO$_2$, the graphene structure constant $\alpha$ is expected to be around $0.5$ and thus $q_{TF}\sim2k_{F}$. In this case, the variation of the extended charge defect potential ($\phi_{\bigtriangleup,\textrm{eff}}$) with the angles $\{\varphi_{i}\}$ is hindered by a TF wave-vector $q_{TF}$ with the same order of magnitude than $Q(\varphi_i)$, as direct inspection of Eq.~(\ref{eq:epl3}) shows. As a consequence, $\sigma_{l}$ becomes little sensitive to the particular random distribution adopted.  
At the Fermi level, the averaged scattering potential reads (see Appendix), 
\begin{equation}
\langle\tilde\phi_{\bigtriangleup,\textrm{eff}}\rangle=\frac{q_{l}}{4\pi k_{F} \epsilon_{d}}\left|\frac{\pi-\theta}{\cos\left(\theta/2\right)}\right|\left[2+e^{ik_{F}(\cos\theta-1)W}\right].
\label{eq:6}
\end{equation}
Interestingly, the line separation $W$ adds an oscillating factor to the scattering amplitude [last term in (\ref{eq:6})], with the consequence that the familiar V-shape in the conductivity as function of the gate voltage \cite{ReviewPeres, RS} (which requires $f(\theta)\sim1/\sqrt{k_{F}}$) will not be observed in samples with a high density of extended charge defects. This manifests into a change of curvature in a $\sigma$ vs $V_{g}$ plot, a \emph{bona fide} signature of scattering due to extended charge defects. This oscillating factor is essential to the high quality fit of the data, as shown in Figs. \ref{fig:fig2} and \ref{fig:fig2b}. 

\subsection{Disorder in the mesh size}
Experimental studies in CVD graphene show extended defects with many geometries, the fact that the $\triangle$ extended defect fits well the data indicates that such defect constitutes the dominant type of scatterer within our model. As discussed above, albeit for $\alpha\sim1/2$ the conductivity is barely affected by the specific orientation of the lines, it is very sensitive to changes in $W$, since this parameter measures roughly the defect mesh size, and hence is directly related to the scattering strength experienced by the electronic carriers. A careful inspection shows that a change of 5$\%$ in $W$ is enough to deteriorate the fits. This parameter sets the departure from the $\sigma \sim n_c$ (or equivalently, $\sigma \sim V$) behavior, similar to that originated from charged impurities located in the substrate, to a more involved carrier-density dependence, namely that of Eq.~(\ref{eq:DC_Conductivity}). 

The effect of disorder in $W$ can be estimated by assuming a normal distribution $p(W)$ with average line separation $\bar{W}$ and variance $dW^{2}$. The potential accounting this kind of disorder is obtained by replacing the exponential factor in Eq.~\ref{eq:6} according to,
\begin{equation}
e^{iq_{x}W} \rightarrow e^{iq_{x}\bar{W}}e^{-\left(q_{x}dW\right)^{2}/2},
\label{eq:epl5b}
\end{equation}
where $q_{x}=k_{F}\left(\cos\theta-1\right)$ is the transferred momentum along the $x$ axis; a finite variance $dW^2$
changes the previous results (i.e. without disorder in $W$) whenever $\left(q_{x}dW\right)^{2}\gtrsim O(1)$. We find $\bar{W}/10\gtrsim dW\gtrsim\bar{W}/100$, in all samples, showing that our model predicts that prominent extended defects form intersection edges with lengths centered around $W\simeq O(10\textrm{nm})$ with small variance.
\subsection{Other scattering mechanisms}
So far we have analyzed the effect of extended charge defects in DC-transport  in graphene. Notwithstanding, other mechanisms are in play which can provide important corrections to our model, especially in the regime of low carrier density, where the fit to Eq.~\ref{eq:DC_Conductivity} is less accurate (see dashed line in Fig.~\ref{fig:fig2}).

We focus our attention on midgap states, presumably the most important scattering mechanism in mechanically cleaved non-suspended graphene samples at finite densities and not too high temperatures \cite{ReviewPeres, geim_new, RS, Monteverde}, here also playing an important role as we will briefly see. Midgap states emerge due to resonant scatterers (RS), whose physical realization could be vacancies or adsorbed hydrocarbons in the surface of graphene. Indeed, we assume a typical value for the density of resonant scatterers, $n_{s}\sim10^{11}\textrm{cm}^{-2}$, and take the radius of the scattering disk $R$ to be of the order of carbon-carbon distance. The correction to the conductivity (Eq.~\ref{eq:DC_Conductivity}) is then calculated using Matthiessen's rule, 
$\sigma^{-1}=\sigma_{l}^{-1}+\sigma_{s}^{-1},$
where, 
\begin{equation}
\sigma_{s}\simeq\frac{2e^2}{h\pi^2 n_{s}}k_{F}^{2}\ln^{2}\left(k_{F}R\right),
\label{eq:epl6}
\end{equation}
is the conductivity due to resonant scatterers \cite{MidGaps, ReviewPeres, RS}. The new fits are obtained by keeping $W$ fixed from its previous value (i.e. with just extended charge defects considered) and varying $\gamma$---Fig.~\ref{fig:fig2} shows that midgap states yield an important correction in the low to moderate density regime.
\subsection{Electron-hole asymmetry}
We finally discuss the origin of the electron-hole asymmetry in DC-transport, highlighted in Fig.~\ref{fig:fig2} by using different colors to represent p and n conductivity. The asymmetry between holes and electrons mobility
\begin{equation}
\mu=(1/e)d\sigma/dn_{c},
\label{eplx}
\end{equation}
is clearly shown in most transport studies in graphene; in CVD samples $|\mu_{p}-\mu_{n}|/\mu_{n}$ is about a few percent. This asymmetry may have two distinct physical origins: (1) a potentially significant
charge transfer from the metallic contacts to graphene \cite{metal_contacts}, and (2) scattering cross-sections sensible to the carriers polarity. 

In order to estimate the contribution of the extended defects to the conductivity asymmetry, we compute the next term in the Born series for a single charged line. The 2$^{\rm nd}$ Born correction to the scattering amplitude $\delta f(\theta)$ is,
\begin{equation*}
\frac{\delta f(\theta)}{f(\theta)}=2p\frac{ev_{F}\hbar}{\tilde\phi_{\textrm{eff}}(\mathbf{q})}\int\frac{d^{2}\mathbf{k}}{\left(2\pi\right)^{2}}\tilde\phi_{\textrm{eff}}(\mathbf{p}^{\prime}-\mathbf{k})G_{D}(\mathbf{k})\tilde\phi_{\textrm{eff}}(\mathbf{k}-\mathbf{p}),
\label{eq:7}
\end{equation*}
where the 2D Dirac propagator reads
\begin{equation}
G_{D}(\mathbf{k})=\frac{1}{v_{F}^{2}\hbar^{2}\left(|\mathbf{p}|^{2}-\mathbf{k}^{2}+i0^{+}\right)},
\label{eq:epl7}
\end{equation}
and the screened potential $\tilde\phi_{\textrm{eff}}$ is given by Eq.~\ref{eq:Potential_Infinite_Line}
with $\lambda_{l}L=q_{l}$. The differential cross-section $|f+\delta f|^{2}$ has now a term proportional to $\tilde\phi_{\textrm{eff}}^3$, and thus is no longer invariant under a change of electrical charge sign $e\rightarrow-e$. For small density of charge $|q_{l}/e|\lesssim0.1$,
the conductivity in the 2$^{\textrm{nd}}$ Born approximation, $\sigma^{(2)}$, relates to the FBA value, $\sigma^{(1)}$, according to 
\begin{equation}
\sigma^{(2)}\simeq\sigma^{(1)}\left[1-g(\alpha) \frac{q_{l}}{e}\right],
\label{eq:epl8a}
\end{equation}
where $g(\alpha)$ reflects the importance of the dielectric medium. For graphene on top of silicon oxide we obtain $g(1/2)\simeq1.2$, entailing a negligible difference 
between the conductivity for carriers with opposite polarities as long as $q_{l}\ll e$. In general, for extended defects made of a single line of charge, the 2$^{\textrm{nd}}$ Born approximation yields a global multiplicative factor in the conductivity; notice that the term inside brackets in Eq.~(\ref{eq:epl8a}) just depends on the charge of the defect and on the sign of carriers, $q_l/e$, and therefore is constant throughout the entire range of carrier density. Indeed, one can absorb the correction from the 2$^{\textrm{nd}}$ Born approximation into $\gamma=n_l q_l/e^2$ according to $\gamma\rightarrow\gamma/[1-g(\alpha)q_l/e]$. 

It would be desirable to perform the same calculation for a $\triangle$ defect. Unfortunately, however, in this case, the 2$^{\textrm{nd}}$ Born correction becomes much involved. Despite that, we can get some intuition by studying a specific configuration; we have performed a numerical evaluation of $g(\alpha)$ for a $N=3$ extended defect with $\varphi_{1(2)(3)}=0$, $\mathbf{R}_{1(2)}=(0,0)$, $\mathbf{R}_{3}=(W,0)$ (see Figure~\ref{fig:fig1} for definitions of $\varphi_{i}$ and $\mathbf{R}_{i}$) and $W=$10nm; a strong dependence of $g(\alpha)$ with $k_{F}$ was observed, more precisely, a variation of $\sim$30$\%$ by increasing the gate voltage from 1V to 50V. We leave as an open question whether the inclusion of the 2$^{\textrm{nd}}$ Born amplitude in the calculation of the conductivity leads to a qualitative improvement of the fits of experimental data to our model --- this would elucidate about the precise amount of electron-asymmetry possibly induced by charged extended in CVD graphene. 

The experimental data shown in Figs.~\ref{fig:fig2} and \ref{fig:fig2b} disclose a small but perceivable change of mobility from p to n carriers. In the light of the latter results, this suggests that the 2$^{\textrm{nd}}$ Born correction cannot be too large in these samples (discarding the remote possibility that a large second order correction is balanced with a charge transfer from the Hall probes). 

\section{Conclusions}

Extended defects are prevalent in CVD graphene and arise in SEM studies as cracks with several sizes and oriented at random. Due to the self-doping mechanism \cite{Peres} such cracks will act as charged scattering centers. In this Letter, we have studied the impact of such defects in the DC-transport properties of graphene films. By constructing a simple model of extended defects, a semi-classical computation of the DC-conductivity was carried out taking into account disorder on the extended defects geometry and the screening by graphene electrons. We have shown that charged extended defects lead to a very distinctive dependence of DC-conductivity with carrier density compared to previously studied mechanisms \cite{ReviewPeres, RS}. Our findings show that extended charge defects can play an important role in DC-transport of chemically grown graphene samples with a large density of cracks. 

Growing graphene via CVD is a very promising route towards scalable fabrication of two-dimensional high-quality carbon films. The understanding of the scattering mechanisms in chemically grown graphene samples is thus of uppermost importance to increase electronic mobilities currently limited to $\simeq$4000 cm$^2 \cdot$V$^{-1} \cdot$s$^{-1}$. Given the stringent constraints on electronic mobilities required for technological applications of graphene, our results show that the control of such defects can be of fundamental importance for further development of a graphene-based electronics. 

\acknowledgments
We thank J. Viana Lopes, V. Pereira and O. V. Yazyev for illuminating discussions. AHCN acknowledges DOE Grant No. DE-FG02-08ER46512 and ONR Grant No. MURI N00014-09-1-1063. AF acknowledges FCT Grant No. SFRH/BPD/65600/2009.

\section{Appendix}

The Thomas-Fermi renormalized electrostatic potential for the extended charge defect $\triangle$ depicted in Fig.~\ref{fig:fig1} reads,
\begin{eqnarray}
\tilde{\phi}_{\bigtriangleup,\textrm{eff}}(\mathbf{q}) & = & \frac{q_{l}}{2\epsilon_{d}}\left[\frac{1}{|q_{x}\cos\varphi_{1}-q_{y}\sin\varphi_{1}|+q_{TF}}\right.\nonumber \\
 &  & +\frac{1}{|q_{x}\cos\varphi_{2}-q_{y}\sin\varphi_{2}|+q_{TF}}\nonumber \\
 &  & \left.+\frac{e^{iq_{x}W}}{|q_{x}\cos\varphi_{3}-q_{y}\sin\varphi_{3}|+q_{TF}}\right].
 \label{eq:scatt_ampl_triangle}
 \end{eqnarray}
The potential depends on the relative orientation of the lines through
the angles $\{\varphi_{i}\}$. To get a meaningful result we have
to consider a statistical distribution of such angles. For $\alpha\gtrsim1/2$
the denominators in (\ref{eq:scatt_ampl_triangle}) are dominated by $q_{TF}$ and hence we can safely integrate the $\{\varphi_{i}\}$ dependences using a uniform distribution. Choosing $\mathbf{q}=k_{F}(\cos\theta-1,\,\sin\theta)$,
we obtain, 
\begin{eqnarray}
\langle\tilde{\phi}_{\bigtriangleup,\textrm{eff}}\rangle & = & \frac{q_{l}}{2\epsilon_{d}k_{F}}\left|\frac{\pi+2\sum_{\beta=-1}^{1}\arctan\left(\frac{f_{\alpha}^{\beta}}{w_{\alpha}}\right)}{\pi w_{\alpha}}\right|\times\nonumber \\
 &  & \times\left(2+e^{ik_{F}\left(\cos\theta-1\right)W}\right),\label{eq:av_pot}\end{eqnarray}
with $w_{\alpha}=\sqrt{2\left(8\alpha^{2}+\cos\theta-1\right)}$,
$f_{\alpha}^{0}=|\sin\theta|$, $f_{\alpha}^{\pm1}=G_{\alpha}^{\pm}/\left(\sin^{2}\frac{\theta}{2}\right)$, where
\begin{equation}
G_{\alpha}^{\pm}=\left[\cos\theta-1\mp4\alpha\right]\left|\sin\frac{\theta}{2}\right|\pm2\alpha\left|\sin\theta\right|.
\label{eq:Gpm}
\end{equation}
Setting $\alpha=1/2$ in (\ref{eq:av_pot}) one obtains Eq.~\ref{eq:6}.

\end{document}